\documentstyle[aps,epsf,preprint]{revtex}

\begin{document}
\def\btt#1{{\tt$\backslash$#1}}
\draft

\title{Mean-field transport theory for the two-flavour NJL model}

\author{Wojciech Florkowski}
\address{H. Niewodnicza\'nski Institute of Nuclear Physics, 
         Radzikowskiego 152, 31 - 342 Krak\'ow, Poland \\
         and Gesellschaft f\"ur Schwerionenforschung GSI, Postfach 110552,
         D-64220 Darmstadt, Germany}

\date{\today}
\draft

\maketitle

\begin{abstract}
By making decomposition of the Wigner function simultaneously in both
the spinor and the isospin spaces we derive a set of kinetic equations
for the quark distribution functions and the spin densities. A detailed 
analysis of the consequences imposed by the chiral invariance on
the form of the transport equations is presented.

\end{abstract}

\pacs{05.60.+w,11.30.Rd,12.39.-x}

\section{Introduction}
\label{sec:intro}

The purpose of this paper is to generalize the already existing formulation
of the mean-field transport theory for the Nambu -- Jona-Lasinio (NJL)
model \cite{NJL0,NJL}. The main aspect of this generalization is an
extension of the simplified one-flavor approach \cite{FHKN96} to
the more realistic two-flavor case. Formally, this is achieved by
applying the technique of decomposition of the Wigner function in both
the spinor and the isospin spaces. In this way we can also extend
and complement some of the earlier calculations done in the framework
of QED \cite{QED} and QHD (quantum hadrondynamics) \cite{QHD}, where the 
Wigner function was decomposed only in the spinor space. 

\medskip In the present approach we take into account all the coefficients 
in the spinor and isospin decomposition. Thus, although our
formulation is restricted to the mean-field approximation it does not
include any further simplifying assumptions concerning the structure
of the Wigner function. This allows us to study the space-time
evolution of the quark distribution functions and the dynamics of
spin.  We include also the possibility of having nonzero pseudoscalar
condensates, which is crucial in studies of the chiral invariance of the
theory.

\medskip We concentrate in more details on the two special cases. The first 
one corresponds to the exact chiral limit, i.e., to the situation when
the current quark masses vanish. On the other hand, the second case
corresponds to the physical situation when the current quark masses
do not vanish and are (slightly) different from each other. Studying
the chiral limit we investigate how the concepts of chiral invariance
can be {\it explicitly} included in the formulation of the transport
theory. In this case we follow the treatment of Ref. \cite{FHKN96}.

\medskip The study of the transport theory for the NJL model becomes
interesting in the context of the ultra-relativistic heavy-ion collisions.  
One expects that these highly energetic processes offer the possibility 
of creation of a short-lived quark-gluon plasma (QGP)
\cite{QM}. It is very likely that during such a deconfinement phase
transition the chiral symmetry is additionally restored. The last fact
is indicated by the lattice simulations of QCD \cite{K90} showing that
the two phase transitions occur at the same temperature. Transport
theory based on the NJL model gives us the possibility to study 
phenomena connected with the chiral phase transition, which take place in
the systems out of thermodynamic and chemical equilibrium. This is an
attractive feature of the model, since the methods for applying QCD
directly to such situations have not been elaborated yet. The QCD
transport theory has been formulated in papers by Heinz \cite{Heinz}
and by Elze {\it et al.} \cite{Elze} but this approach is not
frequent in practical applications.

\medskip In the mean-field approximation the NJL model includes only
the quark degrees of freedom. Therefore, this approach is not fully
appropriate for the description of hadronic matter at low temperatures
or densities and one has to go beyond the mean-field
approximation in order to obtain the agreement of the low-temperature
NJL results with chiral perturbation theory \cite{FB96}. On the
other hand, close to the deconfinement and chiral phase transitions
the quark degrees of freedom become relevant.  In this situation the
mean-field approach is appropriate since it describes many important
features of the chiral phase transition like, e.g., the decrease of
the in-medium quark condensate. Consequently, the present formulation
of the transport theory is most suitable for description of the
phenomena happening in the neighborhood of the chiral phase
transition. Work concerning inclusion of the meson degrees of
freedom into the transport theory for the NJL model (this is 
equivalent to the extension of the mean-field approach) is being
currently carried out by the Heidelberg group \cite{HD0}. 

\medskip The paper is organized as follows. In the next Section we define
the model. In Sect. \ref{sec:wf} we introduce the Wigner function,
define its spinor and isospin decomposition, and discuss the chiral
transformation rules. Sect. \ref{sec:kine} presents the quantum
kinetic equations satisfied by the coefficients of the
spinor and isospin decomposition. In Sect. \ref{sec:con} we do the
classical approximation and derive the so-called constraint equations.
We do it separately for massless and massive quarks. The classical
kinetic equations for the quark distribution functions and for spin
densities are derived in Sect. \ref{sec:classte}. We summarize in Sect.
\ref{sec:sum}.

\section{Definition of the model}
\label{sec:def}

In this paper we take into consideration Lagrangian
\begin{equation}
\label{l}
{\cal L} = {\bar \Psi} \left( i\! \not \! \partial 
- {\hat m} \right) \Psi +
{G \over 2} \left[({\bar \Psi} \Psi)^2 + 
({\bar \Psi} i \gamma_5 \mbox{\boldmath $\tau$} \Psi)^2 \right],
\end{equation}
where $\Psi = (\psi_{u},\psi_{d})$ is the doublet of the Dirac fields,
$G$ is the coupling constant, \mbox{$\mbox{\boldmath $\tau$} = 
(\tau_1,\tau_2,\tau_3)$} are the Pauli isospin matrices, and ${\hat m}$ 
is the matrix containing current quark masses $m_u$ and $m_d$. This matrix
can be written in the form \mbox{${\hat m} = m_0 + {\bf m} \! \cdot \! 
\mbox{\boldmath $\tau$}$}, where \mbox{$m_0 = {\textstyle {1\over 2}} 
(m_u + m_d)$}, $m_1 = m_2 = 0$ and \mbox{$m_3 = {\textstyle {1\over 2}} 
(m_u - m_d)$}. For simplicity, the color degrees of freedom of
quarks have been neglected.

\medskip Using Lagrangian (\ref{l}) and assuming
mean-field approximation we find the following equation for 
the field $\Psi$ 
\begin{equation}
\label{de}
\left[i\! \not\! \partial - {\hat m} - \sigma(x) - i \gamma_5
\mbox{\boldmath $\pi$} (x) \! \cdot \! \mbox{\boldmath $\tau$}\right] 
\Psi(x) = 0.
\end{equation}
Here the mean fields $\sigma(x)$ and $\mbox{\boldmath $\pi$} (x)$ 
are defined through expressions
\begin{equation}
\label{mfas}
\sigma(x) = -G \, \langle \,{\bar \Psi}(x) \Psi(x) \rangle 
= -G \,\, \hbox{Tr}\left( {\hat \rho} \,{\bar \Psi}(x) \Psi(x) \right)
\end{equation}
and
\begin{equation}
\label{mfap}
\mbox{\boldmath $\pi$}(x) = -G \, \langle \,
{\bar \Psi}(x) i \gamma_5 \mbox{\boldmath $\tau$} \Psi(x) \rangle 
= -G \,\, \hbox{Tr}\left( {\hat \rho} \,
{\bar \Psi}(x) i \gamma_5 \mbox{\boldmath $\tau$} \Psi(x) \right),
\end{equation}
where ${\hat \rho}$ is the density operator and $\hbox{Tr}$ denotes
the trace over the Fock space.

\medskip The important feature of Lagrangian (\ref{l}) are different
symmetries which resemble us the case of QCD. First of all Lagrangian
(\ref{l}) is invariant under $\hbox{U}_V(1)$ transformations, which
leads to the conservation of the baryon current
\begin{equation}
\label{uv1}
\partial_{\mu} V^{\mu}(x) = 0, \,\,\,\,\,\,\,
V^{\mu}(x) = \langle {\bar \Psi}(x) \gamma^{\mu} \Psi(x) \rangle.
\end{equation}
In the isospin symmetric case, $m_u = m_d$, Lagrangian (\ref{l}) is 
additionally invariant under $\hbox{SU}_V(2)$ transformations. This
fact gives the conservation of the isospin current
\begin{equation}
\label{suv2}
\partial_{\mu} {\bf V}^{\mu}(x) = 0, \,\,\,\,\,\,\,
{\bf V}^{\mu}(x) = \langle {\bar \Psi}(x) \gamma^{\mu} 
\mbox{\boldmath $\tau$} \Psi(x) \rangle.
\end{equation}
In the chiral limit, $m_u = m_d = 0$, expression (\ref{l}) does not
change under $\hbox{SU}_A(2)$ transformations. This invariance leads
to the conservation of the axial current
\begin{equation}
\label{sua20}
\partial_{\mu} {\bf A}^{\mu}(x) = 0, \,\,\,\,\,\,\,
{\bf A}^{\mu}(x) = \langle {\bar \Psi}(x) \gamma^{\mu} \gamma_5
\mbox{\boldmath $\tau$} \Psi(x) \rangle.
\end{equation}
We note that we have defined the conserved currents as the expectation
values in the state characterized by the density matrix ${\hat \rho}$.
In this case the conservation laws follow directly from Eqs. (\ref{de})
-- (\ref{mfap}).

\medskip Let us now introduce the Green function
\begin{equation}
\label{g}
S^{<}_{i j \, \alpha \beta}(x,y) = \langle {\bar \Psi}_{j \beta}(y)
\Psi_{i \alpha}(x) \rangle,
\end{equation}
where $i,j$ and $\alpha, \beta$ are isospin and spinor indices,
respectively ($i,j = 1,2$ and $\alpha, \beta = 1, ... ,4$).  One can
easily check that the Green function (\ref{g}) fulfills the same
equation as the field $\Psi$ does, namely
\begin{equation}
\label{de1}
\left[i\! \not\! \partial - {\hat m} - \sigma(x) - i \gamma_5
\mbox{\boldmath $\pi$} (x) \! \cdot \! \mbox{\boldmath $\tau$}\right] 
S^{<}(x,y)  = 0.
\end{equation}
Moreover, the mean fields $\sigma(x)$ and $\mbox{\boldmath $\pi$}(x)$
can be determined directly from $S^{<}(x,y)$ via relations
\begin{equation}
\label{sp1}
\sigma(x) = -G \,\, \hbox{tr}\, S^{<}(x,x), \,\,\,\,\,
\mbox{\boldmath $\pi$}(x) = -G \,\, \hbox{tr}\, i\gamma_5 
\mbox{\boldmath $\tau$} S^{<}(x,x), 
\end{equation}
with the trace tr taken over spinor and flavor indices. One can
observe that formulas (\ref{de1}) and (\ref{sp1}) form a closed system
of equations. It will be the subject of our studies in the following
Chapters.

\section{Wigner function}
\label{sec:wf}

\subsection{Decomposition in spinor and isospin space}
\label{subsec:dec}

Our aim is to derive a set of kinetic equations for the classical
distribution functions. This can be achieved in the usual way by
introducing the Wigner function
\begin{equation}
\label{w}
W_{i j \, \alpha \beta}(X,p) = {1 \over \hbar^4}  \int \, d^4u \, 
e^{ {i \over \hbar} p  \cdot  u} \, S^{<}_{i j \, \alpha \beta}
\left(X+ {\textstyle {1\over 2}} u , 
X- {\textstyle {1\over 2}} u \right)
\end{equation}
and by searching for equations fulfilled by this function in the limit
$\hbar \rightarrow 0$. In Eq. (\ref{w}) the quantity $X$ is the 
center-of-mass coordinate, $X = {\textstyle{1\over 2}}(x+y)$, and $u$
is the relative coordinate, $u = x-y$.

\medskip Starting from Eq. (\ref{de1}) and using the well known results 
for the Wigner transform of the derivative of a two-point function,
\mbox{$\partial f(x,y)/\partial x^{\mu}$}, and the Wigner transform of
the product of a one-point function with the two-point one, 
\mbox{$f(x) g(x,y)$}, we find the following equation 
\medskip

\begin{equation}
\label{ke}
\left[ K^{\mu} 
\gamma_{\mu} - M(X) - {\bf m} \!  \cdot  \!  \mbox{\boldmath $\tau$}  
+{i \hbar \over 2} \partial_{\mu} M(X) \, \partial_p^{\mu}- 
i \gamma_5  \, \mbox{\boldmath $\pi$}(X)  \! \cdot  \!  
\mbox{\boldmath $\tau$} 
- {\hbar \over 2} \gamma_5 \, \partial_{\mu}
\mbox{\boldmath $\pi$}(X) \! \cdot \!  \mbox{\boldmath $\tau$} \,
\partial_p^{\mu} \right] W(X,p) = 0,
\end{equation}

\medskip \noindent
where we used the standard notation $\partial^{\mu} = \partial / \partial 
X_{\mu}$, $\partial_p^{\mu} = \partial / \partial p_{\mu}$,
$K^{\mu} =  p^{\mu} + {i \hbar \over 2} \partial^{\mu}$ and
\mbox{$M(X) = \sigma (X) + m_0$}. We note that in Eq. (\ref{ke}) higher
order gradients have been neglected, so it is valid only for
weakly inhomogeneous systems. In addition, there is no collision
term on the right-hand side of Eq. (\ref{ke}), which is a consequence
of our mean-field approximation. The effects of collisions have been
recently studied in Ref. \cite{HD}.

\medskip Since the Wigner function satisfies the condition ${\overline W}(X,p)
= \gamma^0 W^{\dagger}(X,p) \gamma^0 = W(X,p)$, it can be represented
as the following combination of the Dirac tensors $\Gamma$

\begin{equation}
\label{sd}
W = 
{\hat {\cal F}}
+ i\gamma_5 {\hat {\cal P}}
+ \gamma^{\mu} {\hat {\cal V}}_{\mu}
+ \gamma^{\mu} \gamma_5 {\hat {\cal A}}_{\mu} 
+ {1\over 2} \sigma^{\mu \nu}
{\hat {\cal S}}_{\mu \nu}.
\end{equation}
In the decomposition (\ref{sd}) each coefficient ${\hat C}$ [i.e., the
functions ${\hat {\cal F}}(X,p)$,${\hat {\cal P}}(X,p)$,${\hat {\cal
V}}_{\mu}(X,p)$, ${\hat {\cal A}}_{\mu}(X,p)$ and ${\hat {\cal
S}}_{\mu\nu}(X,p)$] is a hermitian two by two matrix. Thus, it can be 
further decomposed in the isospin space according to the rule
\begin{equation}
\label{chat}
{\hat C} = C + {\bf C} \! \cdot \! \mbox{\boldmath $\tau$}.
\end{equation}
In this way we generalize the usual approach, which does not take
into account such structure. Similarly to Eq. (\ref{chat}) we introduce 
the quantity ${\hat M}(X)$ defined as ${\hat M}(X) = M(X) + {\bf m} 
\! \cdot \! \mbox{\boldmath $\tau$} = \sigma(x) + m_0 + {\bf m} 
\! \cdot \! \mbox{\boldmath $\tau$}$. 

\medskip 
Many of the functions defined by the spinor and isospin decomposition,
Eqs. (\ref{sd}) and (\ref{chat}), have a direct physical
interpretation.  For example, the quantities ${\cal V}^{\mu}(X,p),
\mbox{\boldmath ${\cal V}$}^{\mu}(X,p)$ and $\mbox{\boldmath 
${\cal A}$}^{\mu}(X,p)$ are the phase-space densities of the baryon, 
isospin and axial currents. Moreover, according to Eq. (\ref{sp1}), the 
mean fields $\sigma(X)$ and $\mbox{\boldmath $\pi$}(X)$ are simply related 
to the functions ${\cal F}(X,p)$ and $\mbox{\boldmath ${\cal P}$}(X,p)$
\begin{equation}
\label{mfs}
\sigma(X) = -8 G \int {d^4p \over (2\pi)^4}  {\cal F}(X,p) 
\end{equation}
and
\begin{equation}
\label{mfp}
\mbox{\boldmath $\pi$}(X) = 8 G \int {d^4p \over (2\pi)^4}
\mbox{\boldmath ${\cal P}$}(X,p).
\end{equation}

\noindent For the physical interpretation of the other components
we refer the reader to Refs. \cite{QED,BB}.

\subsection{Chiral transformations}
\label{subsec:chii}

The most important feature of Lagrangian (\ref{l}) is its exact chiral
invariance in the case ${\hat m} = 0$. Studying the transport theory
it is interesting to analyze the consequences of this symmetry for the
transport phenomena. This has been already discussed in detail for the
one-flavor version of the model \cite{FHKN96}. In this paper we shall
generalize these results to the two-flavor case.

\medskip The $\hbox{SU}_A(2)$ chiral transformation of the field
$\Psi$ is defined as follows

\begin{equation}
\label{sua2}
\Psi \rightarrow \Psi^{\prime} = \exp(-i\gamma_5 
{ \mbox{\boldmath $\chi$} \! \cdot \! \mbox{\boldmath $\tau$} \over 2}) \Psi 
= \left( \cos {\chi \over 2} - i \gamma_5 {\bf n} \! \cdot \! 
\mbox{\boldmath $\tau$} \sin {\chi \over 2} \right) \Psi .
\end{equation}

\noindent Here ${\bf n}$ is a unit vector in direction of 
$\mbox{\boldmath $\chi$}$ and $\chi$ is the length of 
$\mbox{\boldmath $\chi$}$. The property (\ref{sua2}) induces the 
transformation rules:

\begin{equation}
\label{sigmact}
\sigma  \rightarrow {\sigma^{\prime}} =                 
\sigma \cos \chi -  \mbox{\boldmath $\pi$} \! \cdot \! {\bf n} \sin \chi
\end{equation} 
for the scalar mean field and
\begin{equation}
\label{pionct}
\mbox{\boldmath $\pi$} \rightarrow \mbox{\boldmath $\pi$}^{\prime} =
\mbox{\boldmath $\pi$} \cos^2 {\chi \over 2} - \left[ 2 (
\mbox{\boldmath $\pi$} \! \cdot \! {\bf n} ) {\bf n} - \mbox{\boldmath $\pi$}
\right]  \sin^2 {\chi \over 2} + \sigma {\bf n}  \sin \chi
\end{equation}

\noindent for the pseudoscalar mean field, respectively. As is
expected, Eqs. (\ref{sigmact}) and (\ref{pionct}) indicate that the
sum of squares $\sigma^2 + \mbox{\boldmath $\pi$}^2$ is an invariant of 
the chiral transformations. Another consequence of Eq. (\ref{sua2}) is 
the transformation law for the Wigner function

\begin{equation}
\label{wftr}
W \rightarrow W^{\prime} = \exp(-i\gamma_5 
{ \mbox{\boldmath $\chi$} \! \cdot \! \mbox{\boldmath $\tau$} \over 2})
\,\, W \,\,\exp(-i\gamma_5 
{ \mbox{\boldmath $\chi$} \! \cdot \! \mbox{\boldmath $\tau$} \over 2} ).
\end{equation}

\noindent Eq. (\ref{wftr}) leads to a set of the transformation rules
for the coefficients in the spinor and isospin decomposition.
Their full form is listed in the Appendix. Below we give the
prescriptions for the infinitesimal chiral transformations (denoting
the infinitesimal value of $\mbox{\boldmath $\chi$}$ by $\delta
\mbox{\boldmath $\chi$}$).

\medskip \noindent Infinitesimal chiral transformations of the scalar and 
pseudoscalar coefficients:

\begin{mathletters}
\label{infsps}
\begin{eqnarray}
{\cal F} &\rightarrow& {\cal F}^{\prime} = {\cal F} 
+ \mbox{\boldmath ${\cal P}$} \! \cdot \! {\bf n} \,\, \delta \chi, \\
\mbox{\boldmath ${\cal F}$} &\rightarrow& \mbox{\boldmath ${\cal F}$}^{\prime}
= \mbox{\boldmath ${\cal F}$} + {\cal P} {\bf n} \,\, \delta \chi, \\
{\cal P} &\rightarrow& {\cal P}^{\prime} = {\cal P} 
- \mbox{\boldmath ${\cal F}$} \! \cdot \! {\bf n} \,\, \delta \chi, \\
\mbox{\boldmath ${\cal P}$} &\rightarrow& \mbox{\boldmath ${\cal P}$}^{\prime}
= \mbox{\boldmath ${\cal P}$} - {\cal F} {\bf n} \,\, \delta \chi. 
\end{eqnarray}
\end{mathletters}

\medskip \noindent Infinitesimal chiral transformations of the vector and 
axial-vector coefficients:

\begin{mathletters}
\label{infvav}
\begin{eqnarray}
{\cal V}_{\mu} &\rightarrow& {\cal V}_{\mu}^{\prime} = {\cal V}_{\mu}, \\
\mbox{\boldmath ${\cal V}$}_{\mu} &\rightarrow& 
\mbox{\boldmath ${\cal V}$}_{\mu}^{\prime} = 
\mbox{\boldmath ${\cal V}$}_{\mu}  
- {\bf n} \times \mbox{\boldmath ${\cal A}$}_{\mu} \,\, \delta \chi, \\
{\cal A}_{\mu} &\rightarrow& {\cal A}_{\mu}^{\prime} = {\cal A}_{\mu}, \\
\mbox{\boldmath ${\cal A}$}_{\mu} &\rightarrow& 
\mbox{\boldmath ${\cal A}$}_{\mu}^{\prime} = 
\mbox{\boldmath ${\cal A}$}_{\mu}  
- {\bf n} \times \mbox{\boldmath ${\cal V}$}_{\mu} \,\, \delta \chi.
\end{eqnarray}
\end{mathletters}

\noindent The quantities ${\cal S}_{\mu \nu}, \mbox{\boldmath 
${\cal S}$}_{\mu \nu}, {\tilde {\cal S}}_{\mu \nu}$ and
${\tilde {\mbox{\boldmath ${\cal S}$}}}_{\mu \nu}$ [the last two are
the dual spin tensors defined below in Eq. (\ref{dualst})] transform in
the same way as the functions ${\cal F}, \mbox{\boldmath ${\cal F}$},
{\cal P}$ and $\mbox{\boldmath ${\cal P}$}$. This property resembles
the case of the one-flavor calculation (see Eq. (22) of Ref. \cite{FHKN96}).
In addition, similarly to the one-flavor case one finds that both
${\cal V}^{\mu}$ and ${\cal A}^{\mu}$ are chirally invariant.

\section{Kinetic equations}
\label{sec:kine}

\medskip Substituting formula (\ref{sd}) into Eq. (\ref{ke}) and
comparing the coefficients appearing at the Dirac tensors we find a set 
of coupled equations

\begin{mathletters}
\label{ke1}

\begin{eqnarray}
\label{e1}
K^{\mu} {\hat {\cal V}}_{\mu} - {\hat M} 
{\hat {\cal F}} + 
\mbox{\boldmath $\pi$} \! \cdot \! \mbox{\boldmath $\tau$} 
{\hat {\cal P}} & = &
- {i \hbar \over 2} \left( \partial_{\nu} M\partial_p^{\nu}
{\hat {\cal F}} - \partial_{\nu} \mbox{\boldmath $\pi$} \! \cdot \! 
\mbox{\boldmath $\tau$} \partial_p^{\nu} {\hat {\cal P}} \right), 
\nonumber \\
\\
\label{e2}
-iK^{\mu}{\hat {\cal A}}_{\mu} - {\hat M} 
{\hat {\cal P}} -
\mbox{\boldmath $\pi$} \! \cdot \! \mbox{\boldmath $\tau$} {\hat {\cal F}} 
& = &
- {i \hbar \over 2} \left( \partial_{\nu} M\partial_p^{\nu}
{\hat {\cal P}} + \partial_{\nu} \mbox{\boldmath $\pi$} \! \cdot \! 
\mbox{\boldmath $\tau$} \partial_p^{\nu} {\hat {\cal F}} \right), 
\nonumber \\
\\
\label{e3}
K_{\mu}{\hat {\cal F}} + iK^{\nu}{\hat {\cal S}}_{\nu \mu} 
- {\hat M} {\hat {\cal V}}_{\mu}
+i\mbox{\boldmath $\pi$} \! \cdot \! \mbox{\boldmath $\tau$} 
{\hat {\cal A}}_{\mu} 
&=& -{i \hbar \over 2} \left(\partial_{\nu}
M\partial_p^{\nu}{\hat {\cal V}}_{\mu}
-i\partial_{\nu}\mbox{\boldmath $\pi$} \! \cdot \! 
\mbox{\boldmath $\tau$}\partial_p^{\nu}
{\hat {\cal A}}_{\mu} \right), 
\nonumber \\
\\
\label{e4}
iK^{\mu}{\hat {\cal P}} -K_{\nu} {\hat {\tilde {\cal S}}}^{\nu \mu}
-{\hat M} {\hat {\cal A}}^{\mu} 
+i\mbox{\boldmath $\pi$} \! \cdot \! 
\mbox{\boldmath $\tau$} {\hat {\cal V}}^{\mu} &=& -{i\hbar \over 2} 
\left( \partial_{\nu}
M\partial_p^{\nu}{\hat {\cal A}}^{\mu} -i \partial_{\nu} 
\mbox{\boldmath $\pi$} \! \cdot \! \mbox{\boldmath $\tau$}
\partial_p^{\nu}{\hat {\cal V}}^{\mu} \right), 
\nonumber \\
\\
\label{e5}
\!\!\!i(K^{\mu}{\hat {\cal V}}^{\nu}\!\!-\!K^{\nu}{\hat {\cal V}}^{\mu})
\!-\!\varepsilon^{\mu \nu \alpha \beta} K_{\alpha}{\hat {\cal A}}_{\beta} 
-\! \mbox{\boldmath $\pi$} \! \cdot \! \mbox{\boldmath $\tau$} 
{\hat {\tilde {\cal S}}}^{\mu \nu}
\!+\!{\hat M} {\hat {\cal S}}^{\mu \nu} 
&=&{i \hbar \over 2}
(\partial_{\gamma}M\partial_p^{\gamma}{\hat {\cal S}}^{\mu \nu}\!\!
- \partial_{\gamma}\mbox{\boldmath $\pi$} \! \cdot \! 
\mbox{\boldmath $\tau$} \partial_p^{\gamma}
{\hat {\tilde {\cal S}}}^{\mu \nu}). \nonumber \\
\end{eqnarray}
\end{mathletters}

\noindent In Eqs. (\ref{e4}) and (\ref{e5}) we have introduced the
dual spin tensor ${\hat {\tilde {\cal S}}}^{\mu \nu}$ defined through
relation

\begin{equation}
\label{dualst}
{\hat {\tilde {\cal S}}}^{\mu \nu} = {\textstyle {1 \over 2}}
\varepsilon^{\mu \nu \alpha \beta} {\hat {\cal S}}_{\alpha \beta}.
\end{equation}

\medskip \noindent Equations (\ref{ke1}) represent the {\it spinor} 
decomposition of Eq.  (\ref{ke}). They are a generalization of the set
of equations (24) -- (28) from Ref. \cite{FHKN96} where a one-flavor
formulation has been analyzed. In our two-flavor approach each of
Eqs. (\ref{ke1}) has a matrix form.  In order to obtain a system of
equations that is easier to control, we perform further the {\it
isospin} decomposition. This procedure allows us to deal only with
real quantities, which represent physical observables. At the first
stage of the isospin decomposition we insert expressions of the form
(\ref{chat}) into Eqs.  (\ref{ke1}). Subsequently, we calculate the
sum and the difference of the initial matrix equation and its adjoint
for all the formulas appearing in (\ref{ke1}). In the resulting five
pairs of equations one compares the coefficients at the Pauli isospin
matrices. In doing so, it is convenient to use two relations

\begin{equation}
\label{acom}
{\bf z} \! \cdot \! \{\mbox{\boldmath $\tau$},{\hat C}\} =
2 \, {\bf z} \! \cdot \! {\bf C}+ 2 C \, {\bf z} \! \cdot \! 
\mbox{\boldmath $\tau$}
\end{equation}
and
\begin{equation}
\label{com}
{\bf z} \! \cdot \! [\mbox{\boldmath $\tau$},{\hat C}] = 2 i 
\left({\bf z} \times {\bf C} \right) \mbox{\boldmath $\tau$},
\end{equation}
where ${\bf z}$ is an arbitrary three-vector and $[ , ]$ (\{ , \})
denotes the commutator (anticommutator). Finally, we get five groups
of equations.

\medskip \noindent Scalar equations:

\begin{mathletters}
\label{se}

\begin{eqnarray}
\label{se1}
p^{\mu} {\cal V}_{\mu} 
- M {\cal F} - {\bf m} \! \cdot \!  \mbox{\boldmath ${\cal F}$}
+ \mbox{\boldmath $\pi$} \! \cdot \! 
\mbox{\boldmath ${\cal P}$} & = & 0, \\
\label{se2}
p^{\mu} \, \mbox{\boldmath ${\cal V}$}_{\mu} 
- M \, \mbox{\boldmath ${\cal F}$} - {\bf m} \, {\cal F}  
+ \mbox{\boldmath $\pi$} \, {\cal P} & = & - {\hbar \over 2}
\left(\partial_{\nu} \mbox{\boldmath $\pi$} \times
\partial_p^{\nu} \mbox{\boldmath ${\cal P}$} \right),\\
\label{se3}
{\hbar \over 2} \partial^{\mu} {\cal V}_{\mu} & = & - {\hbar \over 2}
\partial_{\nu} M \partial_p^{\nu} {\cal F} + {\hbar \over 2}
\partial_{\nu} \mbox{\boldmath $\pi$} \! \cdot \! \partial_p^{\nu}
\mbox{\boldmath ${\cal P}$}, \\
\label{se4}
{\hbar \over 2} \partial^{\mu} \mbox{\boldmath ${\cal V}$}_{\mu} 
- {\bf m} \times \mbox{\boldmath ${\cal F}$}
+ \mbox{\boldmath $\pi$} \times \mbox{\boldmath ${\cal P}$}
& = & - {\hbar \over 2} \partial_{\nu} M \partial_p^{\nu}
\mbox{\boldmath ${\cal F}$} + {\hbar \over 2}
\partial_{\nu} \mbox{\boldmath $\pi$} \partial_p^{\nu} {\cal P}.
\end{eqnarray}
\end{mathletters}

\medskip \noindent Pseudoscalar equations:

\begin{mathletters}
\label{pe}
\begin{eqnarray}
\label{pe1}
{\hbar \over 2} \partial^{\mu} {\cal A}_{\mu} - M  {\cal P} -
{\bf m} \! \cdot \! \mbox{\boldmath ${\cal P}$} -
\mbox{\boldmath $\pi$} \! \cdot \! \mbox{\boldmath ${\cal F}$} & = & 0, \\
\label{pe2}
{\hbar \over 2} \partial^{\mu} \mbox{\boldmath ${\cal A}$}_{\mu} - 
M \, \mbox{\boldmath ${\cal P}$} - {\bf m} \, {\cal P} -
\mbox{\boldmath $\pi$} \, {\cal F} & = & {\hbar \over 2}
\partial_{\nu} \mbox{\boldmath $\pi$} \times \partial_p^{\nu}
\mbox{\boldmath ${\cal F}$}, \\
\label{pe3}
p^{\mu} {\cal A}_{\mu} & = & {\hbar \over 2} \partial_{\nu} M 
\partial_p^{\nu} {\cal P} + {\hbar \over 2} \partial_{\nu}
\mbox{\boldmath $\pi$} \! \cdot \! \partial_p^{\nu} 
\mbox{\boldmath ${\cal F}$}, \\
\label{pe4}
{\bf m} \times \mbox{\boldmath ${\cal P}$} +
\mbox{\boldmath $\pi$} \times \mbox{\boldmath ${\cal F}$} 
+ p^{\mu} \mbox{\boldmath ${\cal A}$}_{\mu}
& = &
{\hbar \over 2} \partial_{\nu} M \partial_p^{\nu} 
\mbox{\boldmath ${\cal P}$} + {\hbar \over 2} \partial_{\nu}
\mbox{\boldmath $\pi$} \partial_p^{\nu} {\cal F}.
\end{eqnarray}
\end{mathletters}

\medskip \noindent Vector equations:

\begin{mathletters}
\label{ve}
\begin{eqnarray}
\label{ve1}
p_{\mu} {\cal F} - {\hbar \over 2} \partial^{\nu} {\cal S}_{\nu \mu} 
- M {\cal V}_{\mu} - {\bf m} \! \cdot \! \mbox{\boldmath ${\cal V}$}_{\mu}
& = & - {\hbar \over 2} \partial_{\nu} \mbox{\boldmath $\pi$} \! \cdot \!
\partial_p^{\nu}\mbox{\boldmath ${\cal A}$}_{\mu}, \\
\label{ve2}
- \mbox{\boldmath $\pi$} \times \mbox{\boldmath ${\cal A}$}_{\mu}
+ p_{\mu} \, \mbox{\boldmath ${\cal F}$}
 - {\hbar \over 2} \partial^{\nu} \mbox{\boldmath ${\cal S}$}_{\nu \mu}
- M \, \mbox{\boldmath ${\cal V}$}_{\mu}  - {\bf m} {\cal V}_{\mu} & = &
- {\hbar \over 2} \partial_{\nu} \mbox{\boldmath $\pi$}
\partial_p^{\nu} {\cal A}_{\mu}, \\
\label{ve3}
{\hbar \over 2} \partial_{\mu} {\cal F} + p^{\nu} {\cal S}_{\nu \mu}
+ \mbox{\boldmath $\pi$} \! \cdot \! \mbox{\boldmath ${\cal A}$}_{\mu} & = &
- {\hbar \over 2} \partial_{\nu} M \partial_p^{\nu} {\cal V}_{\mu}, \\
\label{ve4}
- {\bf m} \times \mbox{\boldmath ${\cal V}$}_{\mu} + {\hbar \over 2} 
\partial_{\mu} \mbox{\boldmath ${\cal F}$} + p^{\nu}
\mbox{\boldmath ${\cal S}$}_{\nu \mu} + \mbox{\boldmath $\pi$}
{\cal A}_{\mu} & = & - {\hbar \over 2} \partial_{\nu} M \partial_p^{\nu}
\mbox{\boldmath ${\cal V}$}_{\mu} - {\hbar \over 2} \partial_{\nu}
\mbox{\boldmath $\pi$} \times \partial_p^{\nu}
\mbox{\boldmath ${\cal A}$}_{\mu}.
\end{eqnarray}
\end{mathletters}

\medskip \noindent Axial-vector equations:

\begin{mathletters}
\label{ave}
\begin{eqnarray}
\label{ave1}
- {\hbar \over 2} \partial^{\mu} {\cal P}  - p_{\nu}
{\tilde {\cal S}}^{\nu \mu} - M {\cal A}^{\mu} - {\bf m}
\! \cdot \!  \mbox{\boldmath ${\cal A}$}^{\mu} & = &  - {\hbar \over 2} 
\partial_{\nu} \mbox{\boldmath $\pi$} \! \cdot \! \partial_p^{\nu}
\mbox{\boldmath ${\cal V}$}^{\mu}, \\
\label{ave2}
- \mbox{\boldmath $\pi$} \times \mbox{\boldmath ${\cal V}$}^{\mu}
- {\hbar \over 2} \partial^{\mu} \mbox{\boldmath ${\cal P}$}  - p_{\nu}
{\tilde {\mbox{\boldmath ${\cal S}$}}}^{\nu \mu} 
- M \mbox{\boldmath ${\cal A}$}^{\mu} - {\bf m}
{\cal A}^{\mu} & = & - {\hbar \over 2} 
\partial_{\nu} \mbox{\boldmath $\pi$} \! \cdot \! \partial_p^{\nu}
{\cal V}^{\mu}, \\
\label{ave3}
p^{\mu} {\cal P} - {\hbar \over 2} \partial_{\nu}
{\tilde {\cal S}}^{\nu \mu}  + \mbox{\boldmath $\pi$} \! \cdot \!
\mbox{\boldmath ${\cal V}$}^{\mu} & = & - {\hbar \over 2}
\partial_{\nu} M \, \partial_p^{\nu} {\cal A}^{\mu}, \\
\label{ave4}
- {\bf m } \times \mbox{\boldmath ${\cal A}$}^{\mu} +
p^{\mu} \mbox{\boldmath ${\cal P}$} - {\hbar \over 2} \partial_{\nu}
{\tilde {\mbox{\boldmath ${\cal S}$}}}^{\nu \mu} 
+ \mbox{\boldmath $\pi$} {\cal V}^{\mu} & = & - {\hbar \over 2}
\partial_{\nu} M \, \partial_p^{\nu}
\mbox{\boldmath ${\cal A}$}^{\mu} - {\hbar \over 2}
\partial_{\nu} \mbox{\boldmath $\pi$} \times  \partial_p^{\nu}
\mbox{\boldmath ${\cal V}$}^{\mu}.
\end{eqnarray}
\end{mathletters}

\medskip \noindent Tensor equations:

\begin{mathletters}
\label{te}
\begin{eqnarray}
\label{te1}
- {\hbar \over 2} (\partial^{\mu}{\cal V}^{\nu} - 
\partial^{\nu}{\cal V}^{\mu})
-\varepsilon^{\mu \nu \alpha \beta} p_{\alpha}
{\cal A}_{\beta} - \mbox{\boldmath $\pi$} \! \cdot \!
{\tilde {\mbox{\boldmath ${\cal S}$}}}^{\mu \nu} + M {\cal S}^{\mu \nu} 
+ {\bf m} \! \cdot \! \mbox{\boldmath ${\cal S}$}^{\mu \nu} & = & 0, 
\nonumber \\
\\
\label{te2}
- {\hbar \over 2} (\partial^{\mu}\mbox{\boldmath ${\cal V}$}^{\nu} -
\partial^{\nu}\mbox{\boldmath ${\cal V}$}^{\mu})
-\varepsilon^{\mu \nu \alpha \beta} p_{\alpha}
\mbox{\boldmath ${\cal A}$}_{\beta} - \mbox{\boldmath $\pi$}
{\tilde {\cal S}}^{\mu \nu} + M \mbox{\boldmath ${\cal S}$}^{\mu \nu}
+ {\bf m} {\cal S}^{\mu \nu} & = & {\hbar \over 2} \partial_{\gamma}
\mbox{\boldmath $\pi$} \times \partial_p^{\gamma}
{\tilde {\mbox{\boldmath ${\cal S}$}}}^{\mu \nu} , 
\nonumber \\
\end{eqnarray}
\begin{eqnarray}
\label{te3}
p^{\mu}{\cal V}^{\nu} - p^{\nu}{\cal V}^{\mu} - {\hbar \over 2}
\varepsilon^{\mu \nu \alpha \beta} \partial_{\alpha} {\cal A}_{\beta}
& = & {\hbar \over 2} \left( \partial_{\gamma} M \partial_p^{\gamma}
{\cal S}^{\mu \nu} - \partial_{\gamma} \mbox{\boldmath $\pi$} \! \cdot \!
\partial_p^{\gamma} {\tilde {\mbox{\boldmath ${\cal S}$}}}^{\mu \nu}
\right) , 
\nonumber \\
\\
\label{te4}
- \mbox{\boldmath $\pi$} \times 
{\tilde {\mbox{\boldmath ${\cal S}$}}}^{\mu \nu} + {\bf m} \times
\mbox{\boldmath ${\cal S}$}^{\mu \nu} +
p^{\mu}\mbox{\boldmath ${\cal V}$}^{\nu} -
p^{\nu}\mbox{\boldmath ${\cal V}$}^{\mu}- {\hbar \over 2}
\varepsilon^{\mu \nu \alpha \beta} \partial_{\alpha} 
\mbox{\boldmath ${\cal A}$}_{\beta}
& = & {\hbar \over 2} \left(\partial_{\gamma} M \partial_p^{\gamma}
\mbox{\boldmath ${\cal S}$}^{\mu \nu} - \partial_{\gamma} 
\mbox{\boldmath $\pi$} \partial_p^{\gamma}{\tilde {\cal S}}^{\mu \nu}
\right).
\nonumber \\
\end{eqnarray}
\end{mathletters}

\medskip The first two equations in each set of formulas (\ref{se})
-- (\ref{te}) can be regarded as the ``hermitian'' parts of
Eqs. (\ref{ke1}).  They correspond to Eqs. (31) -- (35) of the
one-flavor approach \cite{FHKN96}. The other equations in formulas (\ref{se}) 
-- (\ref{te}) are the ``anti-hermitian'' parts of Eqs. (\ref{ke1}) and 
correspond to Eqs. (37) -- (41) of \cite{FHKN96}. In the case ${\hat m} = 0$, 
using expressions (\ref{infsps}) and (\ref{infvav}) defining infinitesimal
chiral transformations, one can check that Eqs. (\ref{se}) --
(\ref{te}) are chirally invariant.

One can notice that Eq. (\ref{se3}), integrated over four-momentum $p$,
leads to the baryon current conservation (\ref{uv1}). In the analogous way,
using Eqs. (\ref{se4}) and (\ref{mfp}) one finds the formula
\begin{equation}
\hbar \, \partial_{\mu} {\bf V}^{\mu}(X) = 16 \, {\bf m} \times
\int {d^4p \over (2\pi)^4} \, \mbox{\boldmath ${\cal F}$}(X,p),
\end{equation}
which yields the conservation of the isospin current for the symmetric 
case $m_u = m_d$ [compare Eq. (\ref{suv2})]. Finally, Eqs. (\ref{pe2}),
(\ref{mfs}) and (\ref{mfp}) lead to the expression
\begin{equation}
\label{axial}
\hbar \, \partial_{\mu} {\bf A}^{\mu}(X) = - {2 m_0 \over G} 
\mbox{\boldmath $\pi$}(X) - 16 \, {\bf m} \int {d^4p \over (2\pi)^4}
\, {\cal P}(X,p),
\end{equation}
which is reduced to the axial conservation law, Eq. (\ref{sua20}),
in the chiral limit $m_u = m_d = 0$. Thus we see that after making
the gradient expansion, the conservation laws are still included
in the transport equations.

\section{Constraint equations}
\label{sec:con}

\subsection{Classical Approximation}

In order to obtain classical transport equations one makes an
expansion of the functions ${\hat {\cal F}}(X,p)$, ${\hat {\cal
P}}(X,p)$, ${\hat {\cal V}}_{\mu}(X,p)$, ${\hat {\cal
A}}_{\mu}(X,p)$, ${\hat {\cal S}}_{\mu\nu}(X,p)$, $\sigma(X)$ and
$\mbox{\boldmath $\pi$}(X)$ in powers of $\hbar$. In this way each
coefficient in the decomposition (\ref{sd}) as well as the functions
$\sigma$ and $\mbox{\boldmath $\pi$}$ can be represented as
a series

\begin{equation}
\label{capp}
{\hat C} = {\hat C}_{(0)} + \hbar {\hat C}_{(1)} + 
\hbar^2 {\hat C}_{(2)} + \, ... \,\,\,\,\,\,\, .
\end{equation}

\noindent Inserting expressions of the form (\ref{capp}) into the
kinetic equations (\ref{se}) - (\ref{te}) and comparing the terms 
appearing in the leading (zeroth) order of $\hbar$ we find a set of
constraint equations, which connect different leading order 
terms of the coefficients ${\hat C}$.  

\medskip \noindent Scalar constraint equations:

\begin{mathletters}
\label{cse}

\begin{eqnarray}
\label{cse1}
p^{\mu} {\cal V}^{(0)}_{\mu} 
- M_{(0)} {\cal F}_{(0)} - {\bf m} \! \cdot \!  
\mbox{\boldmath ${\cal F}$}_{(0)} + \mbox{\boldmath $\pi$}_{(0)} \! \cdot \! 
\mbox{\boldmath ${\cal P}$}_{(0)} & = & 0, \\
\label{cse2}
p^{\mu} \, \mbox{\boldmath ${\cal V}$}^{(0)}_{\mu} 
- M_{(0)} \, \mbox{\boldmath ${\cal F}$}_{(0)} - {\bf m} \, {\cal F}_{(0)}  
+ \mbox{\boldmath $\pi$}_{(0)} \, {\cal P}_{(0)} & = & 0, \\
\label{cse3}
- {\bf m} \times \mbox{\boldmath ${\cal F}$}_{(0)}
+ \mbox{\boldmath $\pi$}_{(0)} \times \mbox{\boldmath ${\cal P}$}_{(0)}
& = & 0.
\end{eqnarray}
\end{mathletters}

\medskip \noindent Pseudoscalar constraint equations:

\begin{mathletters}
\label{cpe}
\begin{eqnarray}
\label{cpe1}
M_{(0)}  {\cal P}_{(0)} +
{\bf m} \! \cdot \! \mbox{\boldmath ${\cal P}$}_{(0)} +
\mbox{\boldmath $\pi$}_{(0)} \! \cdot \! 
\mbox{\boldmath ${\cal F}$}_{(0)} & = & 0, \\
\label{cpe2}
M_{(0)} \, \mbox{\boldmath ${\cal P}$}_{(0)} + {\bf m} \, 
{\cal P}_{(0)} +
\mbox{\boldmath $\pi$}_{(0)} \, {\cal F}_{(0)} & = & 0, \\
\label{cpe3}
p^{\mu} {\cal A}^{(0)}_{\mu} & = & 0, \\
\label{cpe4}
{\bf m} \times \mbox{\boldmath ${\cal P}$}_{(0)} +
\mbox{\boldmath $\pi$}_{(0)} \times \mbox{\boldmath ${\cal F}$}_{(0)} 
+ p^{\mu} \mbox{\boldmath ${\cal A}$}^{(0)}_{\mu}
& = & 0.
\end{eqnarray}
\end{mathletters}

\medskip \noindent Vector constraint equations:

\begin{mathletters}
\label{cve}
\begin{eqnarray}
\label{cve1}
p_{\mu} {\cal F}_{(0)} 
- M_{(0)} {\cal V}^{(0)}_{\mu} - {\bf m} \! \cdot \! 
\mbox{\boldmath ${\cal V}$}^{(0)}_{\mu}
& = & 0, \\
\label{cve2}
- \mbox{\boldmath $\pi$}_{(0)} \times \mbox{\boldmath ${\cal A}$}^{(0)}_{\mu}
+ p_{\mu} \, \mbox{\boldmath ${\cal F}$}_{(0)}
- M_{(0)} \, \mbox{\boldmath ${\cal V}$}^{(0)}_{\mu}  
- {\bf m} \, {\cal V}^{(0)}_{\mu} & = & 0, \\
\label{cve3}
p^{\nu} {\cal S}^{(0)}_{\nu \mu}
+ \mbox{\boldmath $\pi$}_{(0)} \! \cdot \! 
\mbox{\boldmath ${\cal A}$}^{(0)}_{\mu} & = & 0, \\
\label{cve4}
- {\bf m} \times \mbox{\boldmath ${\cal V}$}^{(0)}_{\mu} + p^{\nu}
\mbox{\boldmath ${\cal S}$}^{(0)}_{\nu \mu} + \mbox{\boldmath $\pi$}_{(0)}
{\cal A}^{(0)}_{\mu} & = & 0.
\end{eqnarray}
\end{mathletters}

\medskip \noindent Axial-vector constraint equations:

\begin{mathletters}
\label{cave}
\begin{eqnarray}
\label{cave1}
p_{\nu}
{\tilde {\cal S}}_{(0)}^{\nu \mu} + M_{(0)} {\cal A}_{(0)}^{\mu} + {\bf m}
\! \cdot \!  \mbox{\boldmath ${\cal A}$}_{(0)}^{\mu} & = & 0 , \\
\label{cave2}
\mbox{\boldmath $\pi$}_{(0)} \times \mbox{\boldmath ${\cal V}$}_{(0)}^{\mu}
+ p_{\nu} {\tilde {\mbox{\boldmath ${\cal S}$}}}_{(0)}^{\nu \mu} 
+ M_{(0)} \mbox{\boldmath ${\cal A}$}_{(0)}^{\mu} + {\bf m} \,
{\cal A}_{(0)}^{\mu} & = & 0, \\
\label{cave3}
p^{\mu} {\cal P}_{(0)} + \mbox{\boldmath $\pi$}_{(0)} \! \cdot \!
\mbox{\boldmath ${\cal V}$}_{(0)}^{\mu} & = & 0, \\
\label{cave4}
- {\bf m } \times \mbox{\boldmath ${\cal A}$}_{(0)}^{\mu} +
p^{\mu} \mbox{\boldmath ${\cal P}$}_{(0)} + \mbox{\boldmath $\pi$}_{(0)} 
{\cal V}^{\mu}_{(0)} & = & 0.
\end{eqnarray}
\end{mathletters}

\medskip \noindent Tensor constraint equations:

\begin{mathletters}
\label{cte}
\begin{eqnarray}
\label{cte1}
-\varepsilon^{\mu \nu \alpha \beta} p_{\alpha}
{\cal A}^{(0)}_{\beta} - \mbox{\boldmath $\pi$}_{(0)} \! \cdot \!
{\tilde {\mbox{\boldmath ${\cal S}$}}}_{(0)}^{\mu \nu} + M_{(0)} 
{\cal S}_{(0)}^{\mu \nu} 
+ {\bf m} \! \cdot \! \mbox{\boldmath ${\cal S}$}_{(0)}^{\mu \nu} & = & 0, \\
\label{cte2}
-\varepsilon^{\mu \nu \alpha \beta} p_{\alpha}
\mbox{\boldmath ${\cal A}$}^{(0)}_{\beta} - \mbox{\boldmath $\pi$}_{(0)}
{\tilde {\cal S}}_{(0)}^{\mu \nu} + M_{(0)} \mbox{\boldmath 
${\cal S}$}_{(0)}^{\mu \nu}
+ {\bf m} \, {\cal S}^{\mu \nu}_{(0)} & = & 0, \\
\label{cte3}
p^{\mu}{\cal V}_{(0)}^{\nu} - p^{\nu}{\cal V}_{(0)}^{\mu} 
& = & 0, \\
\label{cte4}
- \mbox{\boldmath $\pi$}_{(0)} \times 
{\tilde {\mbox{\boldmath ${\cal S}$}}}_{(0)}^{\mu \nu} + {\bf m} \times
\mbox{\boldmath ${\cal S}$}_{(0)}^{\mu \nu} +
p^{\mu}\mbox{\boldmath ${\cal V}$}_{(0)}^{\nu} -
p^{\nu}\mbox{\boldmath ${\cal V}$}_{(0)}^{\mu}
& = & 0.
\end{eqnarray}
\end{mathletters}

\medskip The constraint equations written above have to be supplemented
by the formulas which determine the values of the mean fields in the
leading order of $\hbar$

\begin{equation}
\label{mfs0}
\sigma_{(0)}(X) = -8G \int {d^4p \over (2\pi)^4} {\cal F}_{(0)}(X,p)
\end{equation}
and
\begin{equation}
\label{mfp0}
\mbox{\boldmath $\pi $}_{(0)}(X) = 8G \int {d^4p \over (2\pi)^4} 
\mbox{\boldmath ${\cal P} $}_{(0)}(X,p).
\end{equation}

\subsection{Chiral limit}

In this subsection we shall analyze the form of the constraint
equations (\ref{cse}) -- (\ref{cte}) in the limit $m_0 = {\bf m} = 0$
[in this case $M_{(0)}(X) = \sigma_{(0)}(X)$].  Having in mind the
previous studies relying on the spinor decomposition
\cite{FHKN96,QED,QHD}, we expect that quantities ${\cal F}_{(0)}$,
$\mbox{\boldmath ${\cal F}$}_{(0)}$, ${\cal A}^{\mu}_{(0)}$ and
$\mbox{\boldmath ${\cal A}$}^{\mu}_{(0)}$ can be used as the
fundamental variables in construction of the transport theory: ${\cal
F}_{(0)}$ describes the quark space-time distribution,
$\mbox{\boldmath ${\cal F}$}_{(0)}$ specifies the isospin of quarks,
${\cal A}^{\mu}_{(0)}$ describes the quark spin density, and
$\mbox{\boldmath ${\cal A}$}^{\mu}_{(0)}$ determines the spin density
of quarks with different isospin components. We note that the
``classical'' isospin of quarks is described here by a
three-vector. This fact is connected with the form of the
decomposition (\ref{chat}) of the two by two, quantum mechanical
density matrix.  We also note that Eqs. (\ref{cpe3}) and (\ref{cpe4})
indicate that only three out of four Lorentz components of ${\cal
A}^{\mu}_{(0)}$ and $\mbox{\boldmath ${\cal A}$}^{\mu}_{(0)}$ are
independent.  Therefore, similarly to the ``classical'' isospin, the
``classical'' spin of quarks is also described by a three-vector.

\medskip Let us now express the quantities ${\cal P}_{(0)}$,
$\mbox{\boldmath ${\cal P}$}_{(0)}$, ${\cal V}_{(0)}^{\mu}$,
$\mbox{\boldmath ${\cal V}$}_{(0)}^{\mu}$, ${\cal S}_{(0)}^{\mu \nu}$,
${\tilde {\cal S}}_{(0)}^{\mu \nu}$, 
$\mbox{\boldmath ${\cal S}$}_{(0)}^{\mu \nu}$ and ${\tilde {\mbox{\boldmath 
${\cal S}$}}}_{(0)}^{\mu \nu}$ as functions of ${\cal F}_{(0)}$, 
$\mbox{\boldmath ${\cal F}$}_{(0)}$, ${\cal A}^{\mu}_{(0)}$ and 
$\mbox{\boldmath ${\cal A}$}^{\mu}_{(0)}$. First of all, it is easy to 
notice that Eqs. (\ref{cpe1}) and  (\ref{cpe2}) define the pseudoscalar 
densities in terms of ${\cal F}_{(0)}$ and $\mbox{\boldmath 
${\cal F}$}_{(0)}$

\begin{equation}
\label{p0}
{\cal P}_{(0)} = 
- { \mbox{\boldmath $\pi$}_{(0)} \! \cdot \! 
\mbox{\boldmath ${\cal F}$}_{(0)} \over \sigma_{(0)} }, 
\end{equation}
\begin{equation}
\label{pv0}
\mbox{\boldmath ${\cal P}$}_{(0)} = 
- { \mbox{\boldmath ${\pi}$}_{(0)}
{\cal F}_{(0)} \over \sigma_{(0)} }.
\end{equation}

\noindent In the similar way Eqs. (\ref{cve1}) and (\ref{cve2}) 
define ${\cal V}^{\mu}_{(0)}$ and $\mbox{\boldmath ${\cal V}$}^{\mu}_{(0)}$
in terms of ${\cal F}_{(0)}$, $\mbox{\boldmath ${\cal F}$}_{(0)}$
and $\mbox{\boldmath ${\cal A}$}^{\mu}_{(0)}$

\begin{equation}
\label{v0}
{\cal V}^{\mu}_{(0)} = p^{\mu} {{\cal F}_{(0)} \over \sigma_{(0)} },
\end{equation}
\begin{equation}
\label{vv0}
\mbox{\boldmath ${\cal V}$}^{\mu}_{(0)}  = p^{\mu}
{ \mbox{\boldmath ${\cal F}$}_{(0)} \over \sigma_{(0)} }
- {\mbox{\boldmath ${\pi}$}_{(0)} \times 
\mbox{\boldmath ${\cal A}$}^{\mu}_{(0)} \over \sigma_{(0)} }.
\end{equation}

\medskip \noindent After a few algebraic manipulations Eqs. (\ref{cte1}) and 
(\ref{cte2}) allow us to express the spin tensors ${\cal S}^{\mu \nu}_{(0)},
\mbox{\boldmath ${\cal S}$}^{\mu \nu}_{(0)}$ and the dual spin tensors 
${\tilde {\cal S}}^{\mu \nu}_{(0)}, {\tilde {\mbox{\boldmath 
${\cal S}$}}}^{\mu \nu}_{(0)}$ as functions of ${\cal A}^{\mu}_{(0)}$ 
and $ \mbox{\boldmath ${\cal A}$}^{\mu}_{(0)}$

\begin{equation}
\label{st0}
{\cal S}^{\mu \nu}_{(0)} = - { 1 \over {\cal M}^2}
\left[p^{\mu} \mbox{\boldmath ${\pi}$}_{(0)} \! \cdot \!
\mbox{\boldmath ${\cal A}$}^{\nu}_{(0)} - p^{\nu}
\mbox{\boldmath ${\pi}$}_{(0)} \! \cdot \!
\mbox{\boldmath ${\cal A}$}^{\mu}_{(0)} \right]
+ { \sigma_{(0)} \over {\cal M}^2}
\varepsilon^{\mu \nu \alpha \beta} p_{\alpha} 
{\cal A}_{\beta}^{(0)},
\end{equation}

\begin{equation}
\label{dst0}
{\tilde {\cal S}}^{\mu \nu}_{(0)} = - { \sigma_{(0)} \over {\cal M}^2}
\left[p^{\mu} {\cal A}^{\nu}_{(0)} - p^{\nu} {\cal A}^{\mu}_{(0)}
\right] - \varepsilon^{\mu \nu \alpha \beta} p_{\alpha}
{\mbox{\boldmath ${\pi}$}_{(0)} \! \cdot \!
\mbox{\boldmath ${\cal A}$}_{\beta}^{(0)} \over {\cal M}^2},
\end{equation}

\begin{equation}
\label{stv0}
\mbox{\boldmath ${\cal S}$}^{\mu \nu}_{(0)} = 
- { \mbox{\boldmath ${\pi}$}_{(0)} \over {\cal M}^2}
\left[p^{\mu} {\cal A}^{\nu}_{(0)} - p^{\nu} {\cal A}^{\mu}_{(0)}
+ \varepsilon^{\mu \nu \alpha \beta} p_{\alpha}
{\mbox{\boldmath ${\pi}$}_{(0)} \! \cdot \!
\mbox{\boldmath ${\cal A}$}_{\beta}^{(0)} \over \sigma_{(0)}} \right]
+ \varepsilon^{\mu \nu \alpha \beta} p_{\alpha}
{\mbox{\boldmath ${\cal A}$}_{\beta}^{(0)} \over \sigma_{(0)}}
\end{equation}

\noindent and

\begin{equation}
\label{dstv0}
{\tilde {\mbox{\boldmath ${\cal S}$}}}^{\mu \nu}_{(0)} = 
{ \mbox{\boldmath ${\pi}$}_{(0)} \over {\cal M}^2}
\left[p^{\mu}  {\mbox{\boldmath ${\pi}$}_{(0)} \! \cdot \! 
\mbox{\boldmath ${\cal A}$}^{\nu}_{(0)} \over  \sigma_{(0)} }
-p^{\nu} {\mbox{\boldmath ${\pi}$}_{(0)} \! \cdot \!
\mbox{\boldmath ${\cal A}$}^{\mu}_{(0)} \over  \sigma_{(0)} } -
\varepsilon^{\mu \nu \alpha \beta} p_{\alpha} {\cal A}_{\beta}^{(0)}
\right]
- {1 \over \sigma_{(0)} } \left( p^{\mu} 
\mbox{\boldmath ${\cal A}$}^{\nu}_{(0)}
- p^{\nu} \mbox{\boldmath ${\cal A}$}^{\mu}_{(0)} \right).
\end{equation}

\bigskip \noindent Here we have introduced the chirally invariant mass 

\begin{equation}
\label{calM}
{\cal M}^2(X) = \sigma_{(0)}^2(X) + \pi_{(0)}^2(X).
\end{equation}

\bigskip Let us now take into consideration other constraint equations
appearing in the leading order of $\hbar$. Substituting expressions 
(\ref{p0}) --  (\ref{vv0}) into Eqs. (\ref{cse1}) and (\ref{cse2}) one 
finds two mass-shell conditions

\begin{equation}
\label{msF}
[p^2 - {\cal M}^2(X) ] {\cal F}_{(0)}(X,p) = 0
\end{equation}
and
\begin{equation}
\label{msvF} 
[p^2 - {\cal M}^2(X) ]\mbox{\boldmath ${\cal F}$}_{(0)}(X,p) = 0.
\end{equation}
The third scalar constraint equation is automatically fulfilled, since 
$\mbox{\boldmath ${\cal P}$}_{(0)}$ and $\mbox{\boldmath ${\pi}$}_{(0)}$
are parallel [see Eqs. (\ref{cse3}) and (\ref{pv0})].

\bigskip Substituting expression (\ref{st0}) in Eq. (\ref{cve3}) and using
condition (\ref{cpe4}) one finds

\begin{equation}
\label{msPivA}
[p^2 - {\cal M}^2(X) ] \, \mbox{\boldmath ${\pi}$}_{(0)} \! \cdot \! 
\mbox{\boldmath ${\cal A}$}^{\mu}_{(0)} = 0.
\end{equation}

\noindent In the analogous way, substituting (\ref{stv0}) in (\ref{cve4}) 
and using condition  (\ref{cpe3}) we get

\begin{equation}
\label{msA}
[p^2 - {\cal M}^2(X) ] {\cal A}^{\mu}_{(0)} = 0.
\end{equation}

\medskip \noindent This formula follows also from Eqs. (\ref{cave1}),
(\ref{dst0}) and (\ref{cpe2}). On the other hand, Eqs. (\ref{cave2}),
(\ref{dstv0}), (\ref{cpe3}) and (\ref{msPivA}) lead to the mass-shell
constraint

\begin{equation}
\label{msvA}
[p^2 - {\cal M}^2(X) ] \, \mbox{\boldmath ${\cal A}$}^{\mu}_{(0)} = 0.
\end{equation}

To complete our discussion of the constraint equations in the chiral
limit we should discuss an important property of formulas
(\ref{mfs0}) and (\ref{mfp0}).  Using Eq. (\ref{pv0}) in (\ref{mfp0})
we find that Eqs. (\ref{mfs0}) and (\ref{mfp0}) are not independent
--- they do not determine separately the values of $\sigma_{(0)}(X)$
and $\mbox{\boldmath $\pi $}_{(0)}(X)$.  This fact is a consequence of
the chiral symmetry. It indicates that only the invariant mass ${\cal
M}(X)$ has a physical significance.

\subsection{Massive quarks}
\label{subsect:mq}

In this subsection we are going to consider the case ${\bf m} \not =
0$. Calculating $\mbox{\boldmath ${\cal P}$}_{(0)}$ from
Eq. (\ref{cpe2}) and substituting into Eq. (\ref{mfp0}) we find that
$\mbox{\boldmath $\pi$}_{(0)}$ and ${\bf m}$ must be parallel in this
case. Using again Eq. (\ref{cpe2}) we find that $\mbox{\boldmath
${\cal P}$}_{(0)}$ must be also parallel to ${\bf m}$.  Further
inspection of Eqs. (\ref{cse}) -- (\ref{cte}) indicates that all
quantities ${\bf C}_{(0)}$ are parallel to ${\bf m}$ [i.e., only their
third component is different from zero, ${\bf C}_{(0)} = (0,0,C_{(0)
\, 3})$] . In this situation it is convenient to introduce the
combinations
\begin{equation}
\label{uad}
C_{(0) \, u} = C_{(0)} + C_{(0) \, 3}, \,\,\,\,\,\, 
C_{(0) \, d} = C_{(0)} - C_{(0) \, 3},
\end{equation}
which describe the up and down quarks, respectively. Using this notation, 
we can write
\begin{equation}
\label{puad}
{\cal P}_{(0)\,u} = -\pi_{(0)\,3} \,\,
{{\cal F}_{(0)\,u} \over M_{(0)} + m_3 },
\,\,\,\,\,\, 
{\cal P}_{(0)\,d} =  \pi_{(0)\,3} \,\,
{{\cal F}_{(0)\,d} \over M_{(0)} - m_3 }
\end{equation}
and
\begin{equation}
\label{vuad}
{\cal V}^{\mu}_{(0)\,u} = {p^{\mu} {\cal F}_{(0)\,u} 
\over M_{(0)} + m_3 },
\,\,\,\,\,\, 
{\cal V}^{\mu}_{(0)\,d} = {p^{\mu} {\cal F}_{(0)\,d} 
\over M_{(0)} - m_3 }.
\end{equation}

\medskip In the massive case the gap equations  (\ref{mfs0}) and 
(\ref{mfp0}) take the form

\begin{equation}
\label{mfsm}
1 + 4G \int {d^4p \over (2\pi)^4 } \left( 
{ {\cal F}_{(0)\,u}(X,p) \over \sigma_{(0)}(X) } +
{ {\cal F}_{(0)\,d}(X,p) \over \sigma_{(0)}(X) } \right) = 0
\end{equation}
and
\begin{equation}
\label{mfpm}
\pi_{(0)\,3}(X) \, \left[
1 + 4G \int {d^4p \over (2\pi)^4 } \left(
{ {\cal F}_{(0)\,u}(X,p) \over \sigma_{(0)}(X) + m_u } +
{ {\cal F}_{(0)\,d}(X,p) \over \sigma_{(0)}(X) + m_d } \right) \right] = 0.
\end{equation}
These two equations lead us to the condition $\pi_{(0)\,3}(X) =
0$. Thus, in the case when the chiral symmetry is explicitly broken,
the leading contribution to the pseudoscalar condensate should vanish
(similarly as in the one-flavor case, see Eq.  (86) of \cite{FHKN96}).

\medskip Substituting expressions (\ref{vuad}) into the scalar constraint
equations (\ref{cse1}) and (\ref{cse2}), we find the mass-shell conditions

\begin{equation}
\label{mcuad}
\left[p^2 - (M^f(X))^2 \right] {\cal F}_{(0)\,f}(X,p) = 0
\hspace{1cm} (f=u,d),
\end{equation}
where we have introduced the notation
\begin{equation}
\label{muad}
M^u(X) = M_{(0)}(X) + m_3, \,\,\,\,\,\,M^d(X) = M_{(0)}(X) -  m_3.
\end{equation}

\medskip 
We turn now to the discussion of the spin dynamics.  First of all,
Eqs. (\ref{cpe3}) and (\ref{cpe4}) give us the condition

\begin{equation}
\label{transc}
p_{\mu} {\cal A}^{\mu}_{(0) \, f} = 0.
\end{equation}
The tensor constraint equations (\ref{cte1}) and (\ref{cte2}) can be
used to find the formulas for the spin tensors, namely

\begin{equation}
\label{stuad}
{\cal S}^{\mu \nu}_{(0) \, f} = {1 \over M^f }
\varepsilon^{\mu \nu \alpha \beta} 
p_{\alpha} {\cal A}_{\beta}^{(0) \, f}.
\end{equation}
The dual spin tensors are obtained by contracting this expression 
with the Levi-Civita tensor as in (\ref{dualst}). Using now Eqs. 
(\ref{transc}), (\ref{stuad}), (\ref{cave1}) and (\ref{cave2}) we
find
\begin{equation}
\label{mcuad1}
\left[p^2 - (M^f(X))^2 \right] {\cal A}^{\mu}_{(0)\,f}(X,p) = 0.
\end{equation}

\section{Classical transport equations}
\label{sec:classte}

In this Section we derive classical transport equations. This requires
the study of Eqs. (\ref{se}) - (\ref{te}) up to the first order in
$\hbar$. Since the calculations done in the chiral limit have different
aspects from those done for massive current quarks, we discuss these
two cases separately.

\subsection{Chiral limit}

Substituting Eqs. (\ref{pv0}) and (\ref{v0}) into Eq. (\ref{se3}) we find

\begin{equation}
\label{kef0}
p^{\mu} \partial_{\mu} F + {\cal M} \partial_{\nu} {\cal M}
\partial_p^{\nu} F = 0,
\end{equation}
where $F(X,p) = {\cal F}_{(0)}(X,p) / \sigma_{(0)}(X)$ is the chirally
invariant quark distribution function (compare Eq. (60) of
\cite{FHKN96}).  In the similar way, substituting (\ref{p0}),
(\ref{vv0}) and the formula for $\mbox{\boldmath ${\cal P} $}(X,p)$
obtained from (\ref{pe2}) into Eq. (\ref{se4}) one gets
\begin{equation}
\label{kef}
p^{\mu} \partial_{\mu} {\bf F} + {\cal M} \partial_{\nu} 
{\cal M} \partial_p^{\nu} {\bf F} = \partial_{\mu} \left(
{ \mbox{\boldmath $\pi $}_{(0)} \over \sigma_{(0)} } \right)
\times \mbox{\boldmath ${\cal A} $}^{\mu}_{(0)},
\end{equation}
where ${\bf F}(X,p) = \mbox{\boldmath ${\cal F}$}_{(0)}(X,p) /
\sigma_{(0)}(X)$. In contrast to $F(X,p)$ the function ${\bf F}(X,p)$ is 
not chirally invariant [under infinitesimal chiral transformations 
${\bf F} \rightarrow {\bf F}^{\prime} = {\bf F} +
\mbox{\boldmath $\pi$}_{(0)} \times \left( {\bf F} \times {\bf n} \right) 
\left( \delta \chi / \sigma_{(0)} \right) $]. However, one can 
check that the {\it form} of Eq. (\ref{kef}) is chirally invariant.

\medskip In order to find the kinetic equations satisfied by the functions
${\cal A}^{\mu}_{(0)}(X,p)$ and 
$\mbox{\boldmath ${\cal A} $}^{\mu}_{(0)}(X,p)$ we use Eqs. (\ref{ave3})
and (\ref{ave4}), respectively. Calculating 
${\cal P}(X,p)$ from Eq. (\ref{pe1}) and 
$\mbox{\boldmath ${\cal V} $}^{\mu}_{(0)}(X,p)$ from Eq. (\ref{ve2}),
and substituting these two expressions into Eq. (\ref{ave3}) we find
(to the first order in $\hbar$)
\begin{equation}
\label{s1}
p^{\mu} \partial_{\nu} {\cal A}^{\nu}_{(0)} +
{\cal M} \partial_{\nu} {\cal M} \partial_p^{\nu}
{\cal A}^{\mu}_{(0)} +
\sigma_{(0)} \partial_{\nu}  {\tilde {\cal S}}^{\mu \nu}_{(0)}
+ \mbox{\boldmath $\pi$}_{(0)} \! \cdot \! \partial_{\nu} 
\mbox{\boldmath ${\cal S}$}^{\mu \nu}_{(0)} = 0.
\end{equation}
In the analogous way, calculating $\mbox{\boldmath ${\cal P}$}(X,p)$ from
Eq. (\ref{pe2}) and ${\cal V}^{\mu}(X,p)$ from Eq. (\ref{ve1}), and 
inserting these expressions into Eq. (\ref{ave4}) we find (again to the
first order in $\hbar$)
\begin{equation}
\label{s2}
p^{\mu} \partial_{\nu} \mbox{\boldmath ${\cal A}$}^{\nu}_{(0)} +
{\cal M} \partial_{\nu} {\cal M} \partial_p^{\nu}
\mbox{\boldmath ${\cal A}$}^{\mu}_{(0)} +
\sigma_{(0)} \partial_{\nu}  
{\tilde {\mbox{\boldmath ${\cal S}$}}}^{\mu \nu}_{(0)}
+ \mbox{\boldmath $\pi$}_{(0)}  \partial_{\nu} 
{\cal S}^{\mu \nu}_{(0)} = 0.
\end{equation}
Using now expressions (\ref{st0}) -- (\ref{dstv0}), defining the
leading order parts of the spin tensors, we find the desired equations
\begin{eqnarray}
\label{s3}
& & p^{\nu} \partial_{\nu} {\cal A}^{\mu}_{(0)}  +
{\cal M} \partial_{\nu} {\cal M} \partial_p^{\nu}
{\cal A}^{\mu}_{(0)} + {\partial_{\nu} {\cal M} \over {\cal M} }
\left[ p^{\mu} {\cal A}^{\nu}_{(0)} - p^{\nu} {\cal A}^{\mu}_{(0)} \right]
\nonumber \\
& & -\varepsilon^{\mu \nu \alpha \beta} p_{\alpha} 
{ {\cal M} \over \sigma_{(0)}}
\partial_{\nu} \left( \mbox{\boldmath $\pi$}_{(0)} \over {\cal M} \right)
\! \cdot \! \mbox{\boldmath ${\cal A}$}_{\beta}^{(0)} = 0
\end{eqnarray}
and
\begin{eqnarray}
\label{s4}
& & p^{\nu} \partial_{\nu} \mbox{\boldmath ${\cal A}$}^{\mu}_{(0)} +
{\cal M} \partial_{\nu} {\cal M} \partial_p^{\nu}
\mbox{\boldmath ${\cal A}$}^{\mu}_{(0)} + {\sigma_{(0)} \over {\cal M}^2}
\partial_{\nu} \left(\mbox{\boldmath $\pi$}_{(0)} \over \sigma_{(0)} \right)
\left[p^{\mu} \mbox{\boldmath $\pi$}_{(0)} \! \cdot \! 
\mbox{\boldmath ${\cal A}$}^{\nu}_{(0)} - p^{\nu} 
\mbox{\boldmath $\pi$}_{(0)} \! \cdot \!
\mbox{\boldmath ${\cal A}$}^{\mu}_{(0)} \right] + 
\nonumber \\
& & {\partial_{\nu} \sigma_{(0)} \over \sigma_{(0)} } \left[p^{\mu}
\mbox{\boldmath ${\cal A}$}^{\nu}_{(0)} - p^{\nu}
\mbox{\boldmath ${\cal A}$}^{\mu}_{(0)} \right]
-\varepsilon^{\mu \nu \alpha \beta} p_{\alpha}
{\sigma_{(0)}^2 \over {\cal M}^2} 
\partial_{\nu} \left( \mbox{\boldmath $\pi$}_{(0)} \over \sigma_{(0)} \right)
{\cal A}^{(0)}_{\beta}
= \sigma_{(0)} \partial^{\mu} \mbox{\boldmath $\pi$}_{(0)} \times
{\bf F}.
\end{eqnarray}

\noindent Several comments are in order now:

\medskip
(a) One can check that Eqs. (\ref{s3}) and (\ref{s4}) are consistent
with the conditions $p^{\mu} {\cal A}_{\mu}^{(0)}= 0$ and $p^{\mu}
\mbox{\boldmath ${\cal A}$}_{\mu}^{(0)} = \mbox{\boldmath ${\cal F}$}_{(0)} 
\times \mbox{\boldmath $\pi$}_{(0)}$. One can also check by a 
straightforward but tedious calculation that Eqs. (\ref{s3}) and
(\ref{s4}) are chirally invariant.

\medskip
(b) The kinetic equations (\ref{kef0}), (\ref{kef}), (\ref{s3}) and
(\ref{s4}) together with the mass-shell constraints (\ref{msF}),
(\ref{msvF}), (\ref{msA}) and (\ref{msvA}) describe the evolution of
quarks in the {\it given} scalar and pseudoscalar fields. The
calculations within the NJL model require that these fields are
obtained self-consistently from the quark distribution functions.
However, the discussed set of equations can be also used in the cases
where we are interested in the dynamics of quarks in the {\it
externally given} scalar and pseudoscalar fields. For that reason, the
results presented in this Section are an extension of a part of the
earlier results obtained in \cite{R95,ZH96}. An interesting novel feature
of the present approach is the coupling between the spin and isospin
degrees of freedom.

\medskip
(c) Doing calculations in the NJL model, one evaluates the mean fields
from the self-consistent equations (\ref{mfs}) and
(\ref{mfp}). However, due to the chiral symmetry of the model [see
discussion following Eqs. (\ref{msvA})] only a chirally invariant
combination ${\cal M}^2 = \mbox{\boldmath $\pi$}_{(0)}^2 +
\sigma_{(0)}^2$ can be calculated from the knowledge of the quark
distribution function. Thus, the discussed system of equations is closed
(and can be solved) only in the case when the ratio $\mbox{\boldmath 
$\pi$}_{(0)}(X)/\sigma_{(0)}(X)$ is fixed and independent of the space-time 
position coordinate $X$. Nevertheless, this ratio can be arbitrary, which is 
required by the chiral invariance of the system. 

\medskip
(d) Since the ratio $\mbox{\boldmath $\pi$}_{(0)}(X)/\sigma_{(0)}(X)$
is fixed and arbitrary, in the practical calculations we can always
set it to zero (this is like choosing the convenient gauge in the
QED-type calculations). In this case the form of the kinetic equations
(\ref{kef}), (\ref{s3}) and (\ref{s4}) simplifies substantially and
reduces to that known from the one-flavor approach (compare Eqs. (62)
and (68) of Ref. \cite{FHKN96}).

\medskip
(e) Eq. (\ref{s4}) does not guarantee the conservation of the axial
current, see Eqs. (\ref{sua20}) and (\ref{axial}). This fact can be
seen in the special case, $\mbox{\boldmath
$\pi$}_{(0)}(X)/\sigma_{(0)}(X)=0$, by using similar arguments to
those presented in the one flavor study \cite{FHKN96}. Consequently,
solutions of Eq.  (\ref{s4}) are {\it constrained} by the condition of
the axial current conservation. Only those solutions which satisfy
Eq. (\ref{sua20}) are valid.

\subsection{Massive quarks}
\label{subsect:kemq}

Using Eq. (\ref{pe2}) and presenting similar arguments to those used
in Section \ref{sec:con}, we find that the cross product
$\mbox{\boldmath $\pi$} \times \mbox{\boldmath ${\cal P}$}$ vanishes
up to the first order in $\hbar$. In this situation Eqs. (\ref{se3})
and (\ref{se4}) lead to the kinetic equations satisfied by the
distribution functions of the up and down quarks. They have the
following form

\begin{equation}
\label{kefq}
p^{\mu} \partial_{\mu} F^f(X,p) + M^f(X) \partial_{\mu} M^f(X)
\partial_p^{\mu} F^f(X,p) = 0 \,\,\,\,\,\, (f=u,d),
\end{equation}
where $F^f(X,p) = {\cal F}_{(0)\, f}(X,p) / M^f(X)$.

\medskip The kinetic equations for the spin densities follow from Eqs. 
(\ref{ave3}) and (\ref{ave4}). Using the formulas for the pseudoscalar
densities ${\cal P}$ and ${\cal P}_3$, calculated from
Eqs. (\ref{pe1}) and (\ref{pe2}) up to the first order in $\hbar$, we
can rewrite (\ref{ave3}) and (\ref{ave4}) in the form

\begin{equation}
\label{sse1}
{\textstyle {1\over 2}} p^{\mu} \partial^{\nu} {\cal A}^{(0)}_{\nu}
-m_3 \left( p^{\mu} {\cal P}_{(1)\, 3} + \pi_{(1)\,3} {\cal V}^{\mu}_{(0)}
\right) + {\textstyle {1\over 2}} M_{(0)} \partial_{\nu} 
{\tilde {\cal S}}^{\mu \nu}_{(0)} = - {\textstyle {1\over 2}} M_{(0)}
 \partial_{\nu}  M_{(0)} \partial_p^{\nu} {\cal A}_{(0)}^{\mu}
\end{equation}
and
\begin{equation}
\label{sse2}
{\textstyle {1\over 2}} p^{\mu} \partial^{\nu} {\cal A}^{(0)\,3}_{\nu}
-m_3 \left( p^{\mu} {\cal P}_{(1)} + \pi_{(1)\,3} {\cal V}^{\mu}_{(0)\,3}
\right) + {\textstyle {1\over 2}} M_{(0)} \partial_{\nu} 
{\tilde {\cal S}}^{\mu \nu}_{(0)\,3} = - {\textstyle {1\over 2}} M_{(0)}
 \partial_{\nu}  M_{(0)} \partial_p^{\nu} {\cal A}_{(0)\,3}^{\mu}.
\end{equation}
Eqs. (\ref{pe1}) and (\ref{pe2}) give us additionally two conditions
\begin{equation}
\label{sse3}
M^u {\cal P}_{(1)\, u} =
{\textstyle {1\over 2}} \partial_{\nu} {\cal A}^{\nu}_{(0)\, u}
-\pi_{(1)\,3} {\cal F}_{(0)\, u}
\end{equation}
and
\begin{equation}
\label{sse4}
M^d {\cal P}_{(1)\, d} =
{\textstyle {1\over 2}} \partial_{\nu} {\cal A}^{\nu}_{(0)\, d}
+\pi_{(1)\,3} {\cal F}_{(0)\, d}.
\end{equation}
Substituting Eqs. (\ref{vuad}), (\ref{sse3}) and (\ref{sse4}) into the
sum of Eqs. (\ref{sse1}) and (\ref{sse2}), and using
formulas for the dual spin tensors we can find
\begin{equation}
\label{sse5}
p^{\nu} \partial_{\nu} {\cal A}_{(0)\,u}^{\mu}
+ M^u \partial_{\nu} M^u \partial_p^{\nu} {\cal A}_{(0)\,u}^{\mu}
+ {\partial_{\nu} M^u \over M^u} \left(p^{\mu}
{\cal A}_{(0)\,u}^{\nu} - p^{\nu} {\cal A}_{(0)\,u}^{\mu} \right) = 0.
\end{equation}
In the similar way, substituting Eqs. (\ref{vuad}), (\ref{sse3}) and 
(\ref{sse4}) into the difference of Eqs. (\ref{sse1}) and (\ref{sse2})
one finds
\begin{equation}
\label{sse6}
p^{\nu} \partial_{\nu} {\cal A}_{(0)\,d}^{\mu}
+ M^d \partial_{\nu} M^d \partial_p^{\nu} {\cal A}_{(0)\,d}^{\mu}
+ {\partial_{\nu} M^d \over M^d} \left(p^{\mu}
{\cal A}_{(0)\,d}^{\nu} - p^{\nu} {\cal A}_{(0)\,d}^{\mu} \right) = 0.
\end{equation}
Equations (\ref{sse5}) and (\ref{sse6}) are the analogs of the 
spin kinetic equation derived for the first time in \cite{FHKN96}.

\bigskip 
The mass-shell conditions allow us to express
$F^{u,d}(X,p)$ as the sum of the quark and antiquark distribution
functions $f^+_{u,d}(X,{\bf p})$ and $f^-_{u,d}(X,{\bf p})$

\begin{equation}
\label{deltas}
F^{u,d}(X,p) = 2\pi \left\{
{\delta\left(p^0-E^{u,d}_p(X)\right) \over 2 E^{u,d}_p(X) }
f^+_{u,d}(X,{\bf p}) + 
{\delta\left(p^0+E^{u,d}_p(X)\right) \over 2 E^{u,d}_p(X) }
\left[f^-_{u,d}(X,-{\bf p}) - 1 \right] \right\}
\end{equation}
where
\begin{equation}
\label{ep}
E^{u,d}_p(X) = \sqrt{(M^{u,d}(X))^2 + {\bf p}^2 }.
\end{equation}
Substituting this formula into Eq. (\ref{kefq}), and integrating over
$p^0$ gives

\begin{equation}
\label{kefq1}
p^{\mu} \partial_{\mu} f^{\pm}_{u,d}(X,{\bf p}) + M^{u,d}(X) 
\partial_{\mu} M^{u,d}(X)
\partial_p^{\mu} f^{\pm}_{u,d}(X,{\bf p}) = 0.
\end{equation}
Using expression (\ref{deltas}) we can also rewrite the gap equation in
the more familiar form 
\begin{eqnarray}
\label{gape}
M_{(0)}-m_0 &=& -2G \int {d^3p \over (2\pi)^3 } {M^u \over E^u_p(X) }
\left[f^+_u(X,{\bf p}) + f^-_u(X,{\bf p}) - 1 \right] \nonumber \\
& & -2G  \int {d^3p \over (2\pi)^3 } {M^d \over E^d_p(X) }
\left[f^+_d(X,{\bf p}) + f^-_d(X,{\bf p}) - 1 \right].
\end{eqnarray}

\medskip
The last results indicate that the form of the kinetic equations for
the two-flavor approach (with massive current quarks) is analogous to
that of the one-flavour model. The up and down quarks, as well as
their spins, evolve in the mean fields $M^u$ and $M^d$,
respectively. This allows to introduce the spin up and spin down
densities as the appropriate combinations of the functions $F^f(X,p)$
and ${\cal A}^{\mu}_{(0) \, f} (X,p)$ (compare Eqs. 69 and 70 of
\cite{FHKN96}).  The only interplay between the two flavors occurs 
via the gap equation, which determines the common part of the mean 
fields $M^u$ and $M^d$.

\section{Summary}
\label{sec:sum}

In this paper we have derived and analyzed the mean-field transport 
equations for the two-flavor NJL model. In this way we have extended
the previous approach \cite{FHKN96}, restricted to only one flavor.
Applying technique of the decomposition of the Wigner function
in both the spinor and isospin space, we could investigate the {\it most
general} form of the mean-field classical transport equations. 

\medskip
We have discussed in detail the case of a chiral limit, studying the
invariance properties of the transport equations. Our analysis shows
the limitations in the applications of the chirally invariant kinetic
equations. This sort of restrictions is already known from the
one-flavor considerations: the ratio $\mbox{\boldmath$\pi$}_{(0)}(X) /
\sigma_{(0)}(X)$ must be constant and the spin dynamics is constrained 
by the axial current conservation. 

\medskip
An additional outcome of our approach are the transport equations for
quark matter moving in the {\it externally given} scalar and
pseudoscalar fields. They follow from our analysis in the case when
the self-consistency required in the evaluation of the meson mean
fields is relaxed. These equations describe non-trivial couplings
between the spin and isospin degrees of freedom.

\medskip
If the current quarks are massive, the difficulties connected with the
requirement of the chiral invariance are not present.  The space-time
evolution of the up and down quark distribution functions is
determined by the mean fields $M^u(X) = M_{(0)}(X) + {\textstyle
{1\over 2}}(m_u - m_d)$ and $M^d(X) = M_{(0)}(X) - {\textstyle {1\over
2}}(m_u - m_d)$, respectively. The common part of the mean fields,
$M_{(0)}(X)$, is obtained from the self-consistent gap equation.

\acknowledgments I am grateful to Joerg H\"ufner for a very warm
hospitality at the Institute of Theoretical Physics of the Heidelberg
University, where this work was initiated. I also thank Bengt Friman
for clarifying discussions and the information about his new
unpublished results.  This work was supported by the KBN Grant
No. 2P03B 080 12, and by the Stiftung f\"ur Deutsch-Polnische
Zusammenarbeit project 1522/94/LN.

\appendix
\section*{}

In this Appendix we list the chiral transformations rules obeyed by the 
coefficients appearing in the spinor and isospin decomposition of the Wigner 
function (\ref{sd}). 

\medskip \noindent Scalar and pseudoscalar functions:

\begin{mathletters}
\begin{eqnarray}
{\cal F} &\rightarrow& {\cal F}^{\prime} = {\cal F} \cos \chi
+ \mbox{\boldmath ${\cal P}$} \! \cdot \! {\bf n} \sin \chi, \\
\mbox{\boldmath ${\cal F}$} &\rightarrow& \mbox{\boldmath ${\cal F}$}^{\prime}
= \mbox{\boldmath ${\cal F}$} \cos^2 {\chi \over 2} 
- \left[ 2 (\mbox{\boldmath ${\cal F}$} \! \cdot \! 
{\bf n} ) {\bf n} - \mbox{\boldmath ${\cal F}$} \right]
\sin^2 {\chi \over 2} + {\cal P} {\bf n} \sin \chi, \\
{\cal P} &\rightarrow& {\cal P}^{\prime} = {\cal P} \cos \chi
- \mbox{\boldmath ${\cal F}$} \! \cdot \! {\bf n} \sin \chi, \\
\mbox{\boldmath ${\cal P}$} &\rightarrow& \mbox{\boldmath ${\cal P}$}^{\prime}
= \mbox{\boldmath ${\cal P}$} \cos^2 {\chi \over 2} 
- \left[ 2 (\mbox{\boldmath ${\cal P}$} \! \cdot \! 
{\bf n} ) {\bf n} - \mbox{\boldmath ${\cal P}$} \right]
\sin^2 {\chi \over 2} - {\cal F} {\bf n} \sin \chi. 
\end{eqnarray}
\end{mathletters}

\medskip \noindent Vector and axial-vector functions:

\begin{mathletters}
\begin{eqnarray}
{\cal V}_{\mu} &\rightarrow& {\cal V}_{\mu}^{\prime} = {\cal V}_{\mu}, \\
\mbox{\boldmath ${\cal V}$}_{\mu} &\rightarrow& 
\mbox{\boldmath ${\cal V}$}_{\mu}^{\prime} = 
\mbox{\boldmath ${\cal V}$}_{\mu} \cos^2 {\chi \over 2} +
\left[ 2 (\mbox{\boldmath ${\cal V}$}_{\mu} \! \cdot \! 
{\bf n} ) {\bf n} - \mbox{\boldmath ${\cal V}$}_{\mu} \right]
- {\bf n} \times \mbox{\boldmath ${\cal A}$}_{\mu} \sin \chi, \\
{\cal A}_{\mu} &\rightarrow& {\cal A}_{\mu}^{\prime} = {\cal A}_{\mu}, \\
\mbox{\boldmath ${\cal A}$}_{\mu} &\rightarrow& 
\mbox{\boldmath ${\cal A}$}_{\mu}^{\prime} = 
\mbox{\boldmath ${\cal A}$}_{\mu} \cos^2 {\chi \over 2} +
\left[ 2 (\mbox{\boldmath ${\cal A}$}_{\mu} \! \cdot \! 
{\bf n} ) {\bf n} - \mbox{\boldmath ${\cal A}$}_{\mu} \right]
- {\bf n} \times \mbox{\boldmath ${\cal V}$}_{\mu} \sin \chi,
\end{eqnarray}
\end{mathletters}

\medskip \noindent Tensor functions:

\begin{mathletters}
\begin{eqnarray}
{\cal S}_{\mu \nu} &\rightarrow& {\cal S}_{\mu \nu}^{\prime} = 
{\cal S}_{\mu \nu} \cos \chi
+ {\tilde {\mbox{\boldmath ${\cal S}$}}}_{\mu \nu} \! \cdot \! 
{\bf n} \sin \chi, \\
\mbox{\boldmath ${\cal S}$}_{\mu \nu} &\rightarrow& 
\mbox{\boldmath ${\cal S}$}^{\prime}_{\mu \nu}
= \mbox{\boldmath ${\cal S}$}_{\mu \nu} \cos^2 {\chi \over 2} 
- \left[ 2 (\mbox{\boldmath ${\cal S}$}_{\mu \nu} \! \cdot \! 
{\bf n} ) {\bf n} - \mbox{\boldmath ${\cal S}$}_{\mu \nu} \right]
\sin^2 {\chi \over 2} + {\tilde {\cal S}}_{\mu \nu} {\bf n} \sin \chi, \\
{\tilde {\cal S}}_{\mu \nu} &\rightarrow& 
{\tilde {\cal S}}_{\mu \nu}^{\prime} = 
{\tilde {\cal S}}_{\mu \nu} \cos \chi
- \mbox{\boldmath ${\cal S}$}_{\mu \nu} \! \cdot \! {\bf n} \sin \chi, \\
{\tilde {\mbox{\boldmath ${\cal S}$}}}_{\mu \nu} &\rightarrow& 
{\tilde {\mbox{\boldmath ${\cal S}$}}}_{\mu \nu}^{\prime}
= {\tilde {\mbox{\boldmath ${\cal S}$}}}_{\mu \nu} \cos^2 {\chi \over 2} 
- \left[ 2 ({\tilde {\mbox{\boldmath ${\cal S}$}}}_{\mu \nu} \! \cdot \! 
{\bf n} ) {\bf n} - {\tilde {\mbox{\boldmath ${\cal S}$}}}_{\mu \nu} \right]
\sin^2 {\chi \over 2} - {\cal S}_{\mu \nu} {\bf n} \sin \chi. 
\end{eqnarray}
\end{mathletters}


\begin{references}

\bibitem{NJL0} Y. Nambu and G. Jona-Lasinio, Phys. Rev. {\bf 122}
(1961) 345; {\bf 124} (1961) 246.

\bibitem{NJL} General reviews of the NJL model: U. Vogel and W. Weise, Prog. 
Part. and Nucl. Phys. {\bf 27} (1991) 195; S.P. Klevansky, Rev. Mod. Phys. 
{\bf 64} (1992) 649; M.K. Volkov, Phys. Part. Nucl. {\bf 24} (1993) 35; 
T. Hatsuda and T. Kunihiro, Phys. Rep. {\bf 247} (1994) 221.

\bibitem{FHKN96} W. Florkowski, J. H\"ufner, S.P. Klevansky and L. Neise,
Ann. Phys. (N.Y.) {\bf 245} (1996) 445

\bibitem{QED} D. Vasak, M. Gyulassy, and H.-T. Elze, Ann. Phys. 
(N.Y.) {\bf 173} (1987) 462.

\bibitem{QHD} H.-T. Elze et al, Mod. Phys. Lett. {\bf A2} (1987) 451.

\bibitem{QM} `` Quark Matter '96, Proc. 12th. International Conference
on Ultra-Relativistic Nucleus-Nucleus Collisions, Heidelberg, Germany,
1996'', Nucl. Phys. {\bf A610}.


\bibitem{K90} See, for example, F. Karsch, in {\it Quark-Gluon Plasma }
(R.C. Hwa, Ed.), World Scientific, Singapore, 1990.

\bibitem{Heinz} U. Heinz, Phys. Rev. Lett. {\bf 51} (1983) 351; Ann.
Phys. (NY) {\bf 161} (1985) 48.

\bibitem{Elze} H.-Th. Elze, M. Gyulassy and D. Vasak, Nucl. Phys.
{\bf 276} (1986) 706; Phys. Lett. {\bf B177} (1986) 402.

\bibitem{FB96} W. Florkowski and W. Broniowski, Phys. Lett. {\bf B386}
(1996) 62.

\bibitem{HD0} S.P. Klevansky, P. Rehberg, A. Ogura and J. H\"ufner,
eprint hep-ph/9701355.

\bibitem{HD} S.P. Klevansky, A. Ogura and J. H\"ufner, Heidelberg
Preprint HD-TVP-97/02; P. Rehberg and  J. H\"ufner, Heidelberg Preprint
HD-TVP-97/03. 

\bibitem{BB} I. Bialynicki-Birula, P. G\'ornicki and J. Rafelski,
Phys. Rev. {\bf D44} (1991) 1825.

\bibitem{R95} G. R. Shin and J. Rafelski, Ann. Phys. (N.Y.) {\bf 243}
(1995) 65.

\bibitem{ZH96} P. Zhuang and U. Heinz, Phys. Rev. {\bf D53} (1996) 2096.


\end{references}
\end{document}